\documentclass[twocolumn,prl,10pt,superscriptaddress,amsfonts]{revtex4-1}

\usepackage{physics}
\usepackage{graphicx}
\usepackage{hyperref} 
\usepackage{xcolor}
\usepackage{soul}
\renewcommand{\figurename}{Figure}
\makeatletter
\renewcommand*{\fnum@figure}{{\normalfont\bfseries \figurename~\thefigure}}
\renewcommand*{\@caption@fignum@sep}{ $|$}
\makeatother

\begin{document}

\title{Silicon qubit fidelities approaching incoherent noise limits via pulse engineering}


\author{C. H. Yang}
\email[]{henry.yang@unsw.edu.au}
\author{K. W. Chan}

\affiliation{Centre for Quantum Computation and Communication Technology, School of Electrical Engineering and Telecommunications, The University of New South Wales, Sydney, NSW 2052, Australia.}

\author{R. Harper}
\affiliation{Centre for Engineered Quantum Systems, School of Physics, The University of Sydney, Sydney, NSW 2006, Australia.}

\author{W. Huang}
\affiliation{Centre for Quantum Computation and Communication Technology, School of Electrical Engineering and Telecommunications, The University of New South Wales, Sydney, NSW 2052, Australia.}

\author{T. Evans}
\affiliation{Centre for Engineered Quantum Systems, School of Physics, The University of Sydney, Sydney, NSW 2006, Australia.}

\author{J. C. C. Hwang}
\author{B. Hensen}
\author{A. Laucht}
\author{T. Tanttu}
\author{F. E. Hudson}
\affiliation{Centre for Quantum Computation and Communication Technology, School of Electrical Engineering and Telecommunications, The University of New South Wales, Sydney, NSW 2052, Australia.}

\author{S. T. Flammia}
\affiliation{Centre for Engineered Quantum Systems, School of Physics, The University of Sydney, Sydney, NSW 2006, Australia.}

\author{K. M. Itoh}
\affiliation{School of Fundamental Science and Technology, Keio University, 3-14-1 Hiyoshi, Kohoku-ku, Yokohama 223-8522, Japan}

\author{A. Morello}
\affiliation{Centre for Quantum Computation and Communication Technology, School of Electrical Engineering and Telecommunications, The University of New South Wales, Sydney, NSW 2052, Australia.}

\author{S. D. Bartlett}
\email[]{stephen.bartlett@sydney.edu.au}
\affiliation{Centre for Engineered Quantum Systems, School of Physics, The University of Sydney, Sydney, NSW 2006, Australia.}

\author{A. S. Dzurak}
\email[]{a.dzurak@unsw.edu.au}
\affiliation{Centre for Quantum Computation and Communication Technology, School of Electrical Engineering and Telecommunications, The University of New South Wales, Sydney, NSW 2052, Australia.}


\date{\today}

\begin{abstract}

\end{abstract}


\maketitle


%

\textbf{The performance requirements for fault-tolerant quantum computing are very stringent.  Qubits must be manipulated, coupled, and measured with error rates well below 1\%\cite{Knill2005,Fowler2012}.
For semiconductor implementations, silicon quantum dot spin qubits have demonstrated average single-qubit Clifford gate error rates that approach this threshold~\cite{veldhorst2014,Kawakami18102016,Watson2018,Zajac2017}, notably with error rates of 0.14\% in isotopically enriched $^{28}$Si/SiGe devices~\cite{Yoneda2017}.
This gate performance, together with high-fidelity two-qubit gates and measurements, is only known to meet the threshold for fault-tolerant quantum computing in some architectures when assuming that the noise is incoherent, and still lower error rates are needed to reduce overhead.
Here we experimentally show that pulse engineering techniques, widely used in magnetic resonance~\cite{Khaneja2005}, improve average Clifford gate error rates for silicon quantum dot spin qubits to 0.043\%,a factor of 3 improvement on previous best results for silicon quantum dot devices~\cite{Yoneda2017}. 
By including tomographically complete measurements in randomised benchmarking, we infer a higher-order feature of the noise called the unitarity, which measures the coherence of noise. This in turn allows us to theoretically predict that average gate error rates as low as 0.026\% may be achievable with further pulse improvements.
These fidelities are ultimately limited by Markovian noise, which we attribute to charge noise emanating from the silicon device structure itself, or the environment.}

Randomised benchmarking~\cite{Knill2008, Magesan2011,Magesan2012a,wallman2017} is the gold standard for quantifying the performance of quantum gates, and can be used to efficiently obtain accurate estimates of the average gate fidelity in the high-accuracy regime independent of state preparation and measurement (SPAM) errors.  The standard method for randomised benchmarking, however, is designed to provide \emph{only} the average gate fidelity, and not any further details about the noise.  To improve quantum gates further, one would also like diagnostic information about the character of the noise processes, i.e., its frequency spectrum, whether it is primarily due to environmental couplings or control errors, etc.  Quantum tomography methods can provide such information but are in general inefficient and highly sensitive to SPAM errors.  For these reasons, variants of randomised benchmarking that quantify higher-order noise features as well as the average gate fidelity have been developed~\cite{Wallman2015a,Kimmel2014,feng2016}.

As an early example of this approach, the randomised benchmarking data of Ref.~\cite{veldhorst2014} demonstrating average Clifford gate fidelities of 99.59\% in SiMOS qubits exhibited non-exponential decay features, which was subsequently attributed to low-frequency detuning noise in the system~\cite{fogarty2015}.  That is, randomised benchmarking of this device not only demonstrated its high performance, but also provided details of the noise characteristics.  These details in turn suggest a method to further reduce the infidelity:  Low frequency noise is particularly amenable to pulse engineering techniques, which exploit the quasi-static nature of the noise process, and so appropriate engineering of gate pulses should in principle lead to higher fidelities.

\begin{figure}
	\includegraphics{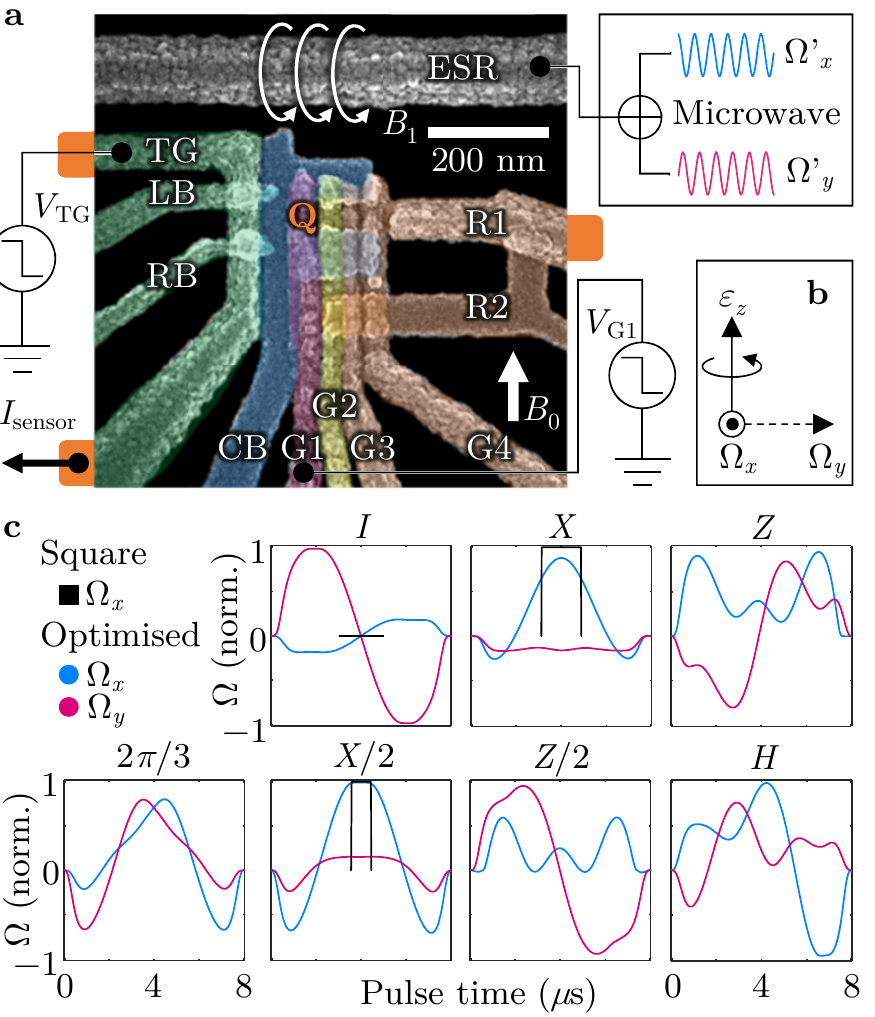}%
	\caption{\label{fig:device}
		\textbf{Device image, experimental setup, and GRAPE optimised Clifford gates.} 
		\textbf{a}, Scanning electron micrograph (SEM) image of a SiMOS qubit device with the same design studied here. 
		The quantum dot that holds our qubit is confined via gate CB laterally and under gate G1, denoted by the letter Q, and has a size of roughly 40nm in diameter. An IQ-modulated microwave source is connected to the ESR line, driving the control field $B_1$. $V_\text{TG}$ and $V_\text{G1}$ are controlled/pulsed during the experiment, and spin-to-charge conversion is detected via $I_\text{sensor}$.
		\textbf{b}, Axes corresponding to terms of the Hamiltonian acting on the qubit in a rotating frame referenced to the microwave, where $H=\Omega_x\sigma_x + \Omega_y\sigma_y + \epsilon_z\sigma_z$. The qubit sees $B_1$ as effective $\Omega_x$ and $\Omega_y$ control axes. Additional noise of $\epsilon_z$ acts on the direction of the DC field $B_0$.
		\textbf{c}, Microwave modulation $\Omega_x$ (blue), $\Omega_y$ (red) for the basic 7 types of Clifford gate ($I, X, Z, 2\pi/3, X/2, Z/2, H$), found through GRAPE iteration. Standard square pulse (black) for $I, X, X/2$ that are used to construct the standard Clifford sequence. All other gates can constructed by having phase shift on the microwave, for example $Y$-phase gates are performed by having a $\pi/2$ phase offset.
		}
		\hspace*{\fill}
	
\end{figure}

Here, we demonstrate that pulse engineering techniques can be used to increase the average Clifford gate fidelity of single-qubit gate operations on the same SiMOS quantum dot, from 99.83\% to 99.96\%. In terms of coherence this leads to an improvement in $T_2^{\text{RB}}$ from $620\mu$s to $9.4$ms, a factor of 15 times improvement compared with square pulses.  See \autoref{fig:device} for details of our qubit experiment, and the shaped optimised pulse used in this work. 
We exploit recent developments in randomised benchmarking to give high-precision, high-credibility estimates of this average gate fidelity.

\begin{figure}
	\includegraphics{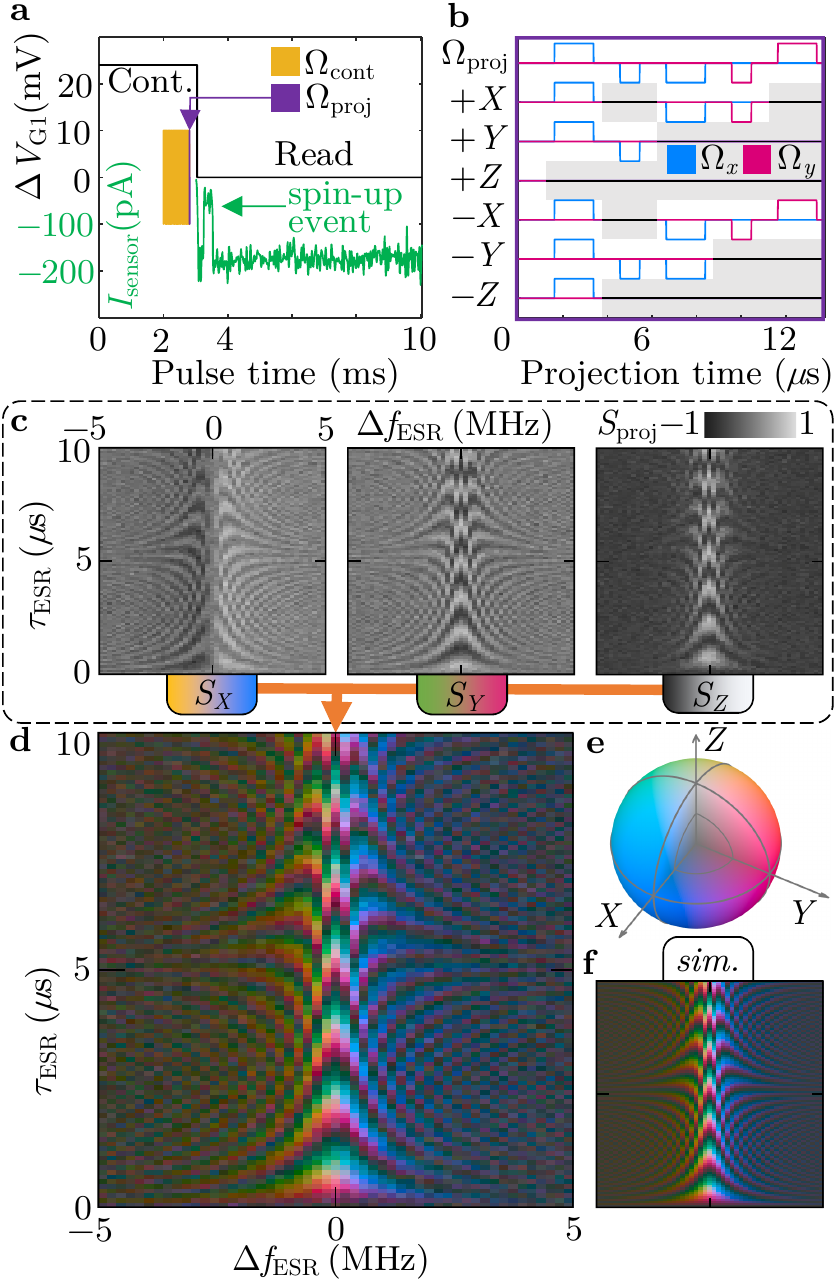}%
	\caption{\label{fig:density}
		\textbf{Density matrix reconstruction through tomographic readout.}
		\textbf{a}, The single shot sequencing of tomographic readout. Immediately after the control pulse sequence finishes, the projection pulse follows, cycling through spin projection axes of $X\Rightarrow Y\Rightarrow Z\Rightarrow-X\Rightarrow-Y\Rightarrow-Z\Rightarrow X \ldots$ 
		\textbf{b}, Microwave modulation for the 6 axis readout pulses. $\Omega_\text{all}$ is the master projection pulse modulation, where the other six axis readout pulses are generated via mask multiplexing (grey shaded) of microwave output.
		\textbf{c}, $X Y Z$ projection of a Rabi oscillation Chevron map, measured and maintained with feedback control (\autoref{fig:feedback}) over 14 hours.
		\textbf{d}, Coloured density matrix of (c). Colour scale of (f) is used.
		\textbf{e}, CIELAB colour space coded Bloch sphere, with prime axis colour: $\pm X$ ($-b^*$ channel, blue-yellow), $\pm Y$ ($a^*$ channel red-green), $\pm Z$ ($L$ channel, white-black), the centre point colour grey represents a fully dephased state. (Colour may saturate due to conversion to RGB)
		\textbf{f}, Simulation of Chevron map in (d), with 80\% readout visibility, with exact $0.8\mu$s~$\pi$-pulse time, no fitting attempted.
		\hspace*{\fill}
	}
\end{figure}

Furthermore, by using tomographic measurements in our randomised benchmarking, we are able to quantify the \emph{unitarity}~\cite{Wallman2015a} of the noise: a higher-order noise feature that  quantifies the average change in purity of a state, averaged over a given gate set, see \autoref{fig:density}.
Inaccurate but highly pure dynamics still have high unitarity. Unitarity allows us to quantify the coherence in the noise independent of the error rate.
Measuring both the unitarity and the average error rate allows us to estimate how much of an experimental error budget is due to control errors and low-frequency noise versus uncorrectable decoherence. 
Our measurements of the unitarity demonstrate that the improved gates from pulse engineering have primarily reduced the unitary component of the noise, while also quantifying the potential future gains to be made from further pulse engineering and control improvements.

\paragraph{Pulse engineering.}
Our previous study of the cause of qubit gate infidelity for SiMOS qubits~\cite{fogarty2015} identified low-frequency drift in the qubit detuning as the dominant noise term.  The timescale of this drift process is very long compared with the timescale for control of the qubit, enabling pulse engineering techniques to be used to identify compensating pulses for this noise term.  Specifically, we use Gradient Ascent Pulse Engineering (GRAPE)~\cite{Khaneja2005} to identify pulses for our qubit control that are robust against low-frequency detuning noise.  This method uses a theoretical model for the noise, together with gradient ascent methods to identify (locally) optimal pulses. 

We identify 7 improved Clifford gate operators using this procedure, as detailed in the supplementary information and \autoref{fig:device}c. The full set of 24 single-qubit Clifford gates can be simply achieved by phase shifting one of the 7 basic operators (manipulating the sign of $\Omega_x$ and $\Omega_y$, and/or swapping them). For example, a $Y$ gate can be constructed via swapping $\Omega_x$ and $\Omega_y$ of the $X$ gate.  

\paragraph{Calibration.}
Four different controllers were used to ensure the spin qubit environment and control parameters did not drift during the whole of the 35 hours-long experiment. (See supplementary information section for full details). Calibration data suggests that the main source of noise $\epsilon_z$ comes from the nuclear-spin from residue 29Si, where the change of resonance frequency has strong step like behaviour and no clear correlation with charge rearrangement. Similar nuclear-spin like behaviour has been observed in the same device while operating in two-qubit mode \cite{Huang2018}.

\paragraph{Tomographic measurement.}
\autoref{fig:density}a shows an example of a single shot sequence during the randomised benchmarking.
The projection pulses that measure each  spin projection are shown in \autoref{fig:density}b. They are designed in a way that can be easily multiplexed and have built-in echoing ability. To further confirm these projection pulses are able to correctly construct a robust density matrix, a tomographic Rabi Chevron map is performed in \autoref{fig:density}c,d, combined with the calibration technique described above.
The colour coded density matrix map packs all the spin projection maps into one, giving a stronger feel for the correlation between $XYZ$ axes.
Surprisingly, the measured and without-fit-simulated data in \autoref{fig:density}f looks almost identical, save that the simulated map has less background white noise. Notice even at far detuning ($\Delta{}f_{\text{ESR}}>3$MHz), where a normal $Z$ projection only Rabi map would appear to have no readout signal, we can still see $XY$ phase oscillation for both sets of data. The coloured Rabi Chevron measurement serves as strong validation of our tomographic readout, feedback control and microwave calibration, and confirms the data quality in our randomised benchmarking experiment.

\paragraph{Randomised benchmarking.}
We assess the performance of our improved gates using randomised benchmarking~\cite{Knill2008, Magesan2011,Magesan2012a}.  Randomised benchmarking and its variants are fully scalable protocols that allow for the partial characterization of quantum devices.  
Here we use a variant of randomised benchmarking to determine the average gate fidelity as well as the coherence (unitarity) of the noise~\cite{Wallman2015a}.  
An overview of randomised benchmarking is given in the methods.

\begin{figure}

	\includegraphics{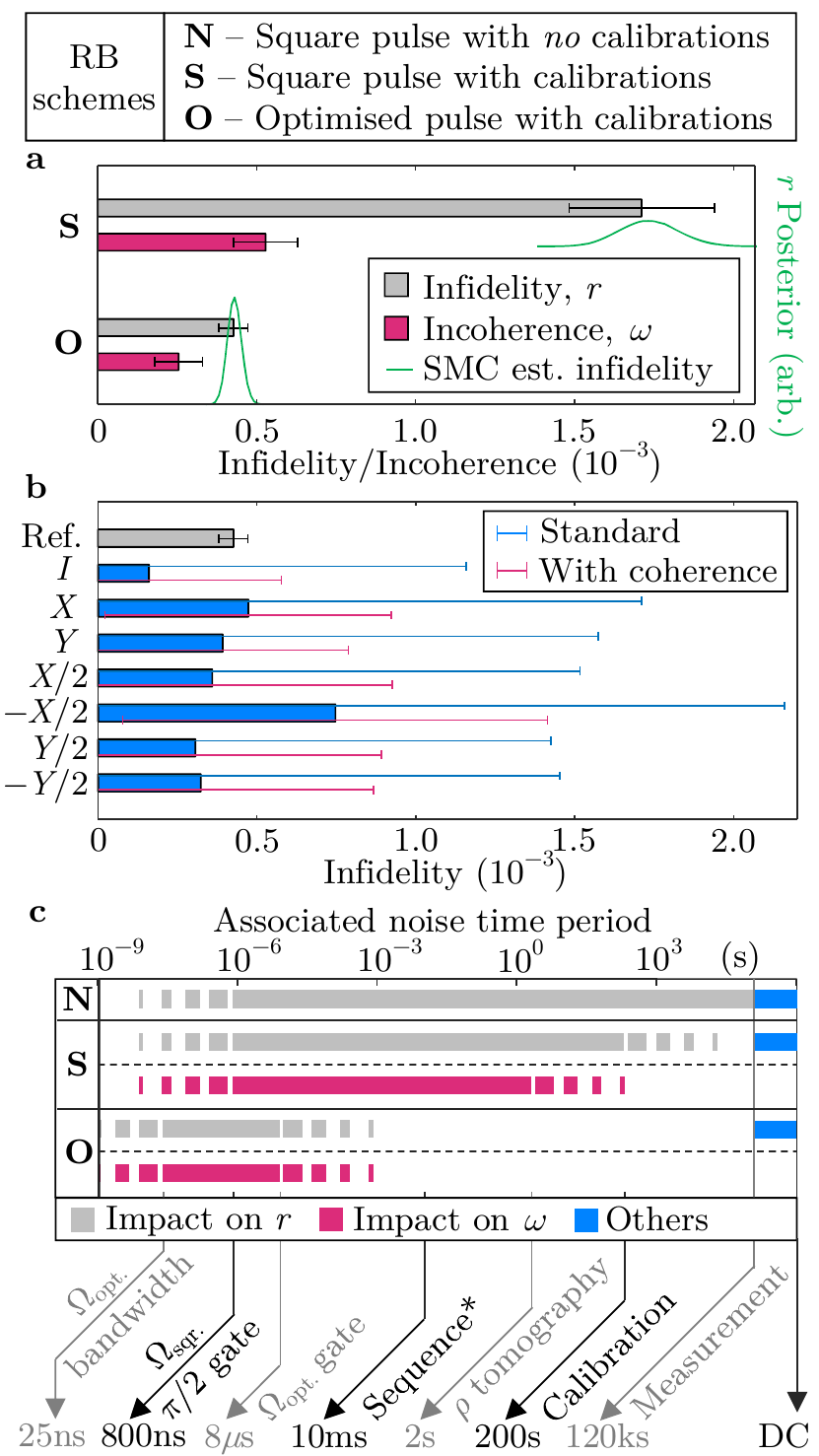}
	\caption{\label{fig:sequences}
		\textbf{Randomised benchmarking result and noise profile}
		\textbf{a}, Infidelity ($1-F$) and incoherence ($\omega$) (see \autoref{eq:omega}) for sequences using square pulses (scheme \textbf{S}) and GRAPE optimised pulses (scheme \textbf{O}), both with calibrations activated. Error bars are calculated using weighted non-linear least squares. The value of~$\omega$ (here in red) indicates the amount of infidelity (grey) that is attributable to incoherent noise. Perfect unitary control should allow the infidelity to be reduced to the value of $\omega$. The green lines are Sequential Monte Carlo estimates of the pulse fidelities, which show the tight credible region on the estimate of the average fidelity for the reference optimised pulse data set.
		\textbf{b}, Analysis of the interleaved gates for scheme \textbf{O}. The error bars to the side of the interleaved gate sequences are calculated using the original method of Ref.~\cite{Magesan2012} and then using an improved method incorporating unitarity~\cite{dugas2016}.
		\textbf{c}, Noise frequency impact on types of randomised benchmarking schemes.
		With scheme \textbf{N}, standard square pulses only (no calibration, not performed in this work), any noise frequency starting from its sub-gate time to DC would affect the system. 
		For scheme \textbf{S}, the calibration should reduce the effect of noise slower than the calibration period. 
		With tomographic readout, the measurement of incoherence will also remain unaffected by noise slower than the tomography time.
		For scheme \textbf{O}, both fidelity and coherence will have reduced impact from noise slower than the actual gate time, but may be more affected from higher frequency noise ---  up to its shaped-pulse bandwidth.
		Other DC errors will have direct impact on fidelity, less so on incoherence. 
		\hspace*{\fill}
	}
\end{figure}

The results of our randomised benchmarking experiments, determining the average gate fidelities of both the original (square) pulses scheme \textbf{S}, and the improved pulses scheme \textbf{O}, are given in \autoref{fig:sequences}. 
Both pulse schemes are performed for each of the measurement projections in an alternating manner, using the identical square projection pulses shown in \autoref{fig:density}b, with calibrations activated.
For scheme \textbf{S}, this gave a measured randomised benchmarking decay factor ($p$) of $99.66(5)\%$, which equates to an  average per-Clifford fidelity of $99.83(2)\%$. With scheme \textbf{O}, this resulted in a decay factor of $99.914(9)\%$, which equates to an average per-Clifford fidelity of $99.957(4)\%$, where the error indicates the 95\% confidence levels. For comparison purposes we note that the literature often reports not only a Clifford gate fidelity but also a fidelity based on gate generators. Here we report only the fidelity returned by randomised benchmarking, namely the \textit{per Clifford fidelity}.  The relevant comparison fidelities are therefore the $99.96\%$ achieved here, compared to $99.86\%$ \cite{Yoneda2017}, $99.90\%$\cite{Muhonen2015} and $99.24\%$ \cite{fogarty2015,veldhorst2014}.  Fitting assumptions and methods are detailed in the supplementary information. Bayesian analysis was carried out leading to the tight credible regions seen in \autoref{fig:sequences}a.

\paragraph{Unitarity.}
The data obtained from the tomographic measurements in our randomised benchmarking experiments allows us to determine the \textit{unitarity}, which is a higher order feature of the noise afflicting the system~\cite{Wallman2015a}.  The unitarity can be used to distinguish `unitary' errors (which may arise from such things as control errors and/or low frequency noise) from stochastic errors (which are generally associated with high-frequency noise).

For a system of dimension $d$, the unitarity is defined as an integral of pure states ($\psi$) over the Haar measure as follows:
\begin{equation}\label{eq:unitarity}
u(\mathcal{E})=\frac{d}{d-1}\int d\psi \Tr\qty[\mathcal{E}\qty(\op{\psi}-\frac{1}{d}\mathbb{I})^2]\,,
\end{equation}
and provides a measure as to where the noise sits between being completely coherent noise, where the unitarity achieves its maximum value of 1, to being completely depolarising noise, where it obtains its minimum value. The minimum value depends on the fidelity and we have $u(\mathcal{E})\geq \bigl[1-\frac{dr}{\qty(d-1)}\bigr]^2$, which is saturated by a completely depolarising channel with average infidelity, $r = 1-F$. 
We can define a new quantity, the \textit{incoherence}, which is related to the unitarity as follows:
\begin{equation}
\omega(\mathcal{E})=\frac{d-1}{d}\left(1-\sqrt{u(\mathcal{E})}\right)\,.\label{eq:omega}
\end{equation}
The incoherence is defined so that it takes a maximum value given by the infidelity, and a minimum value of 0 (purely coherent noise), so that $0\le \omega(\mathcal{E}) \le r(\mathcal{E})$. 
The value of the incoherence  represents the minimum infidelity that might be achievable if one had perfect unitary control over the system. 
Defined as above, the incoherence (when compared to the infidelity) directly gives an indication of the amount of the infidelity that is attributable to incoherent (statistical) noise sources. 

The incoherence therefore allows us to: a) estimate useful information relating to the type of noise afflicting the system; b) provides a guide as to how much improvement can be made to the fidelity of the system by the correction of purely coherent errors (such as over-rotations); and c) can be used to provide tighter bounds on the likely diamond distance of the average noise channel~\cite{Wallman2015,Kueng2016} and to reduce uncertainty in the interleaved benchmarking protocol~\cite{dugas2016}, see Fig.~\ref{fig:sequences}b.

The incoherence (\autoref{eq:omega}) for scheme \textbf{S} is $0.5(1)\times 10^{-3}$ and for the scheme \textbf{O} is $0.2(1)\times 10^{-3}$. Using the incoherence allows a direct comparison with the reported infidelities, see Fig.~\ref{fig:sequences}a. With scheme \textbf{S} the incoherence is approximately $30\%$ of the infidelity and for the improved pulses $61\%$ of the infidelity. Two conclusions can be drawn from this.  First, the data provides strong, quantitative evidence that the improved pulses have reduced the errors on the gates primarily by reducing coherent errors.  Second, we observe that the infidelity for the optimised pulses is below the incoherence for the square pulses, and that there are still coherent errors in the improved gates.  Therefore, by using the scheme \textbf{O} we have not only improved our unitary control but have also reduced the incoherent noise in the system. 

Figure \ref{fig:sequences}c illustrates an intuitive explanation of these results. Scheme \textbf{O} minimises the effect of noise on timescales greater than $8\mu{}s$, decreasing the infidelity of the system and the coherence of the remaining infidelity. The small trade off, however, is that the pulse optimised gates are slightly more susceptible to  higher  frequency noise, up to the bandwidth of the pulse, leaving us with some coherent noise. Finally, near DC imperfections such as miscalibrations and microwave phase errors will also contribute to degrade the fidelity, but with lesser impact on its incoherence.

We have demonstrated that the unitarity can be used not just to characterise noise but also as a tool to increase gate fidelities. As a final note, we observe that the incoherence ($\omega$) is indicative of the infidelity that would in principle be achieved if all coherent errors in the system were eliminated. Specifically, the data indicates that with the improved pulses, if perfect unitary control could be achieved, then the per Clifford fidelity of the gates could be as high as 99.974\%.

\begin{acknowledgments}
We acknowledge support from the US Army Research Office (W911NF-13-1-0024, W911NF-14-1-0098, W911NF-14-1-0103, and W911NF-17-1-0198), the Australian Research Council (CE11E0001017 and CE170100009), and the NSW Node of the Australian National Fabrication Facility. 
The views and conclusions contained in this document are those of the authors and should not be interpreted as representing the official policies, either expressed or implied, of the Army Research Office or the U.S. Government. 
The U.S. Government is authorised to reproduce and distribute reprints for Government purposes notwithstanding any copyright notation herein. B.H. acknowledges support from the Netherlands Organization for Scientific Research (NWO) through a Rubicon Grant. K.M.I. acknowledges support from a Grant-in-Aid for Scientific Research by MEXT, NanoQuine, FIRST, and the JSPS Core-to-Core Program.
\end{acknowledgments}

\clearpage
\section{Supplementary Information}

\subsection{Stochastic Gradient Ascent Pulse Engineering}

We model our qubit system using the Hamiltonian $H=\Omega_x\sigma_x + \Omega_y\sigma_y + \epsilon_z\sigma_z$, where $\Omega_x/\Omega_y$ are the $I$-quadrature/$Q$-quadrature (in-phase/out-phase) microwave amplitudes, and $\epsilon_z$ is a fixed (DC) random variable representing the $Z$ detuning for a single pulse sequence.

The amplitudes $\Omega_x$ and $\Omega_y$, as functions of time, are the two controls available that define our shaped microwave pulse.  In each iteration of GRAPE, we calculate the derivative $\frac{\delta\Psi}{\delta\Omega}$ of the target operator fidelity $\Psi$ corresponding to each sample point $\Omega_x$ and $\Omega_y$, and update them accordingly to maximise $\Psi$.
Our GRAPE implementation is stochastic, sampling $\epsilon_z$ on every iteration from a Gaussian distribution of $\frac{1}{2T_2^*} = 16.7\text{kHz}$ noise strength, where $T_2^*=30\mu s$.  (Note that $\epsilon_z$ is constant within a single iteration.)  In our search for improved pulses, we constrain the maximum pulse length to $8$ $\mu$s, four times longer than a square $\pi$-pulse.  The amplitude of each pulse is also constrained by $\Omega_x^2$ + $\Omega_y^2 = \Omega_\text{max}^2$, where $\Omega_\text{max} = \frac{1}{2T_\pi} = \frac{1}{2\times1.75\mu\text{s}} = 285.7\text{kHz}$ is the maximum allowed effective $B_1$ amplitude. 

Each optimised pulse is constructed via 800 $\Omega$ samples at a sample rate of 10ns, with a time length of 8$\mu$s.

For a given, small, learning factor $\eta$, a single iteration step can be written as:
\begin{enumerate}
	\item Randomise $\epsilon_z$
	\item Calculate $\frac{\delta\Psi}{\delta\Omega}$ for all $\Omega$ point-wise, with the current Hamiltonian $H$
	\item Update $\Omega \rightarrow \Omega + \eta\frac{\delta\Psi}{\delta\Omega}$
	\item Filter $\Omega$ for smoothness and bound condition $\Omega^2_\text{max}\geq\Omega^2_x+\Omega^2_y$
\end{enumerate}

The pulse optimisation can perform roughly 100 iterations per second with MATLAB, within a few minutes solutions that have close infidelities to \autoref{fig:device}(c) can be found. Here, we optimise 7 basic Clifford gate operators using the GRAPE method described above. These basic gates can be expanded to the complete group of 24 Clifford gates by  phase shifting one of the 7 basic operators (manipulating the sign of $\Omega_x$ and $\Omega_y$, and/or swapping them). For example, a $Y$ gate can be constructed via swapping $\Omega_x$ and $\Omega_y$ of the $X$ gate. \autoref{fig:device}(c)a shows the optimised Clifford gates that were found and used for the randomised benchmarking experiment. The normal square pulses in black are plotted in the same scale for comparison.

\begin{figure*}
	\includegraphics{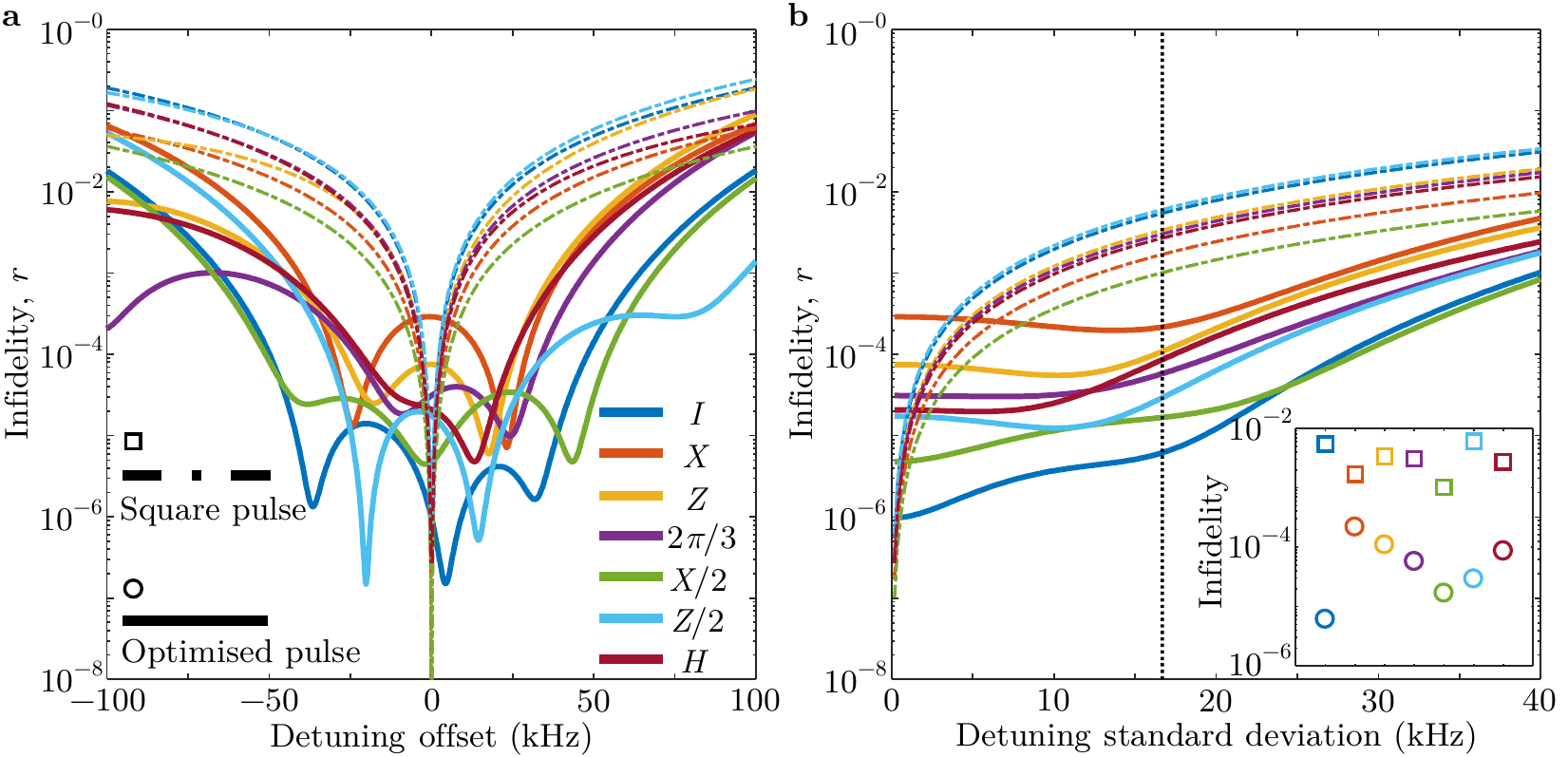}%
	\caption{\label{fig:grapesim}
		\textbf{Simulation for Clifford gates.} 
		\textbf{a}, Infidelity simulation of Clifford gates under fixed detuning, $\sigma_z$ noise.
		\textbf{b}, Infidelity simulation with standard deviation on detuning, $\sigma_z$ noise. 
		The infidelities are calculated through integrating Gaussian type noise on detuning, with mean of~0.
		Inset: The fidelity of Clifford gates where detuning standard deviation is 16.7kHz (dotted line in (c)), the assumed noise power when running GRAPE iteration.
		\hspace*{\fill}
	}
\end{figure*}

\autoref{fig:grapesim}a  shows the  simulation of the infidelities, comparing the optimised pulses and square-pulses for the basic 7 Clifford gates, given a fixed $\sigma_z$ detuning noise. The square pulse constructed Clifford gates have no errors when the detuning noise is 0, but infidelities grow rapidly with increasing detuning. On the other hand, gates constructed from the optimised pulses have higher infidelity when no detuning exists, but can tolerate a wider detuning offset range. \autoref{fig:grapesim}b takes account of the complete $\sigma_z$ noise distribution, and shows the calculated overall infidelity against the $\sigma_z$ standard deviation. Again, if the qubit system has $\sigma_z$ noise level near 0, the square pulse gates would have much lower infidelities. However, when $\sigma_z$ is sitting around a noise level of 16.7kHz, the infidelities are minimised for the optimised case. This pre-experimental simulation shows the pulse optimised Clifford gates have an order of magnitude improvement over the square pulse Clifford gates under the condition of a single noise source $\sigma_z$ being quasi-static with a standard deviation of 16.7kHz. Composite pulses have also been shown to improve fidelity in ion trap experiments~\cite{Mount2015}.

\begin{figure*}
	\includegraphics{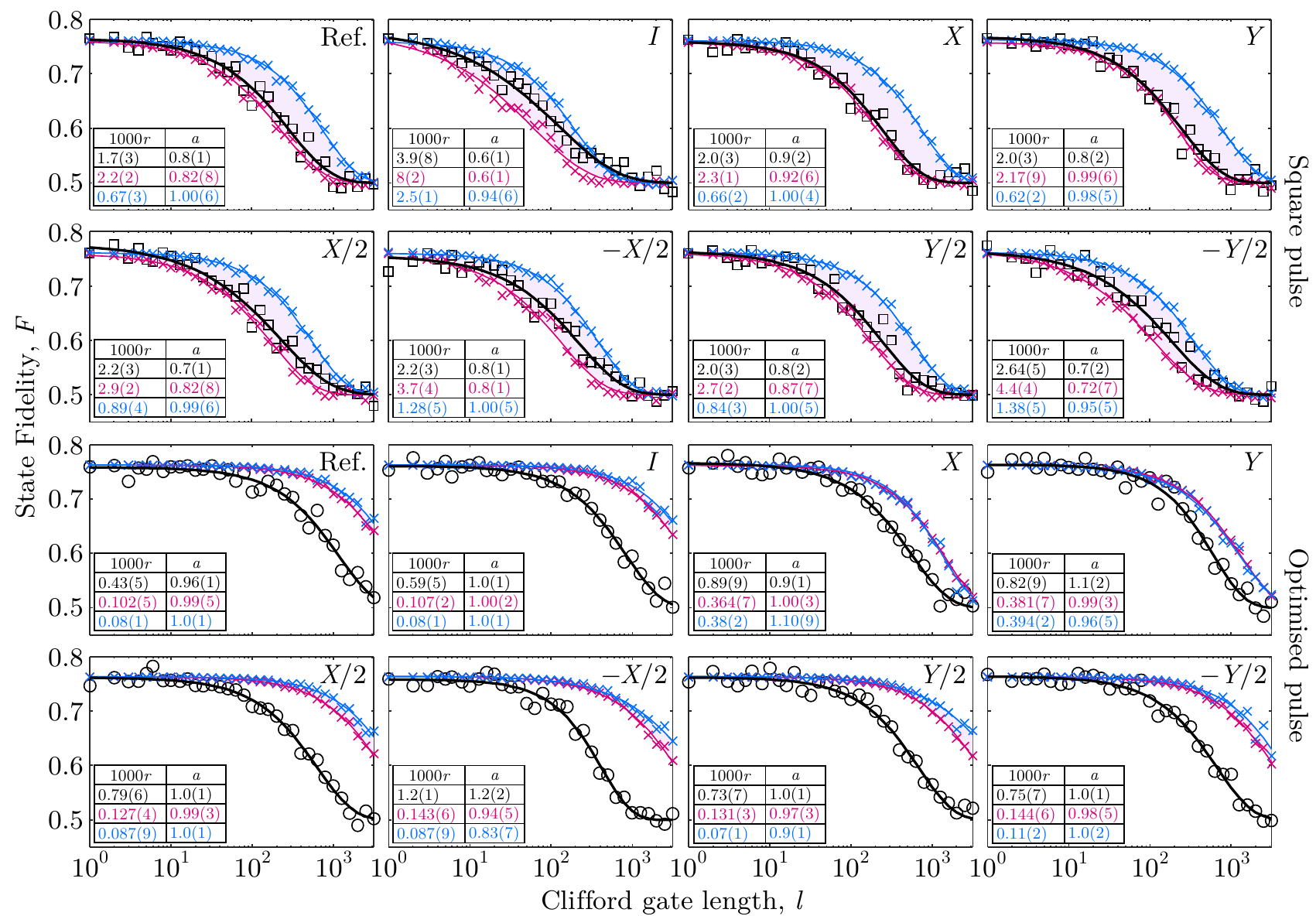}%
	\caption{\label{fig:RBMsim}
		\textbf{Randomised Benchmarking re-simulation} 
		Comparison of experimental data from randomised benchmarking and re-simulation using identical control sequences, with different interleaved gates and pulse schemes. The noise model used in the simulation is a mimic of the GRAPE model, assuming a quasi-static detuning noise, which is updated every 400$\mu$s ($T_{2 \text{Hahn}}$ time). Black trace: Experimental data. Red trace: Re-simulation with detuning noise level of 16.7kHz. Blue trace: Re-simulation with detuning noise level of 7.5kHz. The fitting equation is $F = \frac{A}{2}\exp{-(2\times l\times r)^a}+\frac{1}{2}$, where $F$ is the state fidelity, $l$ is the Clifford gate length, $r$ is the randomised benchmarking infidelity, $a$ is the decay exponent, and $A$ is the visibility.
		\hspace*{\fill}
	}
\end{figure*}

To validate that the noise model also fits the randomised benchmarking result, we re-simulate the experiment using the identical ESR control sequence (identical Clifford sequence) that is being applied for every single shot; see \autoref{fig:RBMsim}. 
The noise is modelled in the same way as the GRAPE model, namely, quasi-static detuning noise with standard deviation of 16.7kHz, which will be updated every 400$\mu$s ($T_{2 \text{Hahn}}$ time). 
An additional noise level of $\frac{1}{\sqrt{2}\pi{}T_2^*} = 7.5\text{kHz}$ is also simulated, which is the Gaussian noise level assuming the $T_2^*$ measurement is infinitely long. 
The close agreement between experimental data and simulation in the square pulses scheme suggests that the noise model introduced in the search for GRAPE pulses fits well. 
However, the expected optimised pulses' performance is much higher than the experimental results. 
This is due to a breakdown of the noise model when higher frequency noise on detuning and microwave control start to dominate when the quasi-static noise is suppressed. 
Noting the low decay exponents in the square pulses scheme (in both experiment and simulation), it again suggests that non-Markovian behaviour arises in randomised benchmarking in the presence of low frequency noise~\cite{fogarty2015}.


\subsection{Feedback and Calibration}\label{sec:feedback}

\begin{figure*}
	\includegraphics{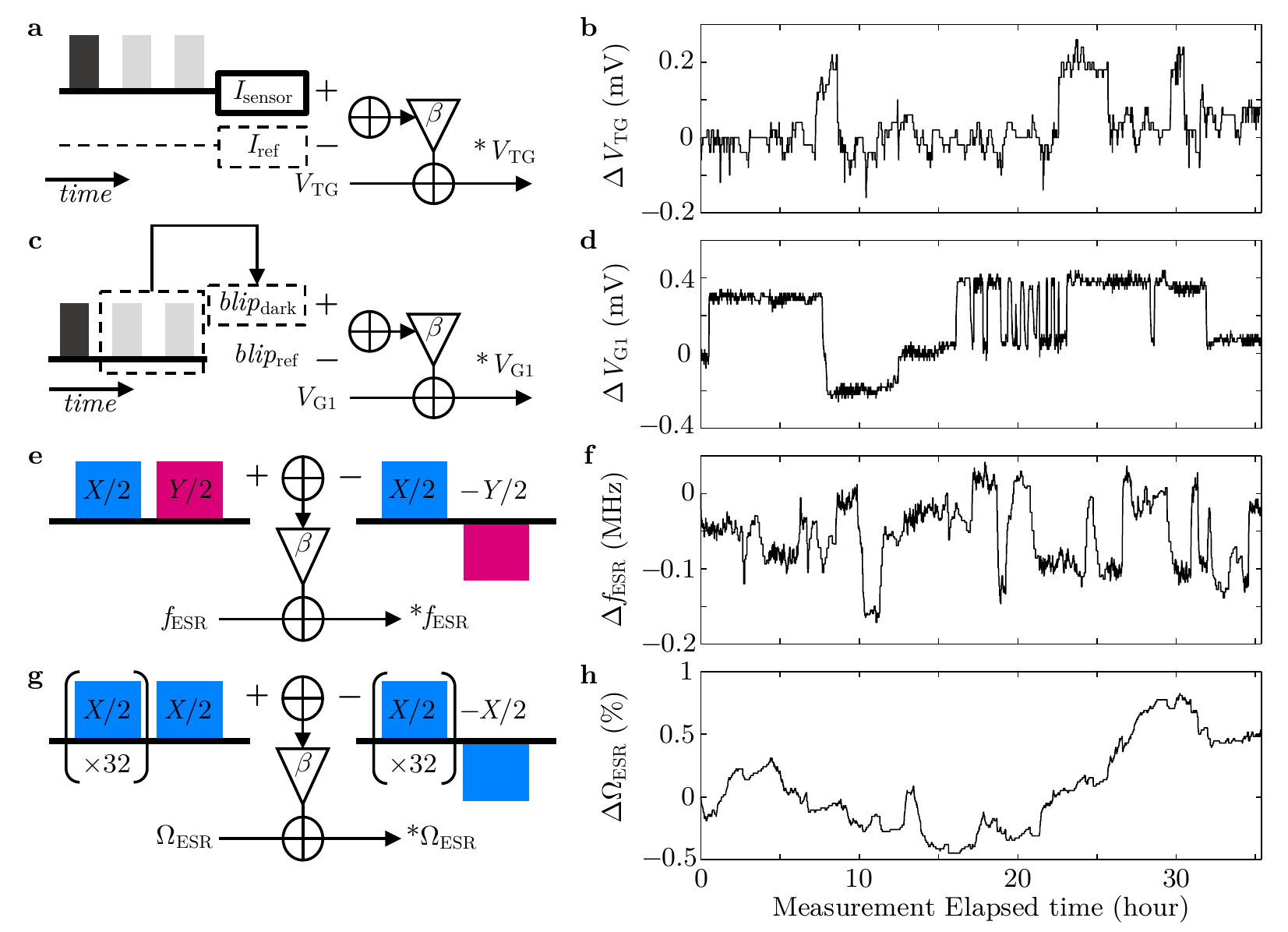}%
	\caption{\label{fig:feedback}
		\textbf{Feedback control and calibration for randomised benchmarking over 35 hours.}
	\textbf{a}, Feedback control of SET sensor bias current. Difference of the SET current and the desired bias current is fed back into $V_\text{TG}$, with a gain of $\beta$. 
	\textbf{b}, Change of $V_\text{TG}$ over the whole randomised benchmarking measurement period. 
	\textbf{c}, Feedback control of spin to charge readout level.
	The dark blip counts (blips that occur at the later stage of the readout window) are maintained at a particular rate via changing $V_\text{G1}$, ensuring the readout visibility is constant. 
	\textbf{d}, Change of $V_\text{G1}$ over the whole randomised benchmarking measurement period. 
	\textbf{e}, Calibration of resonance frequency. The resonance frequency is tracked by taking the difference of two ESR pulse sequences, one does a $X/2$ followed by $Y/2$, and the other with $X/2$ followed by $-Y/2$. The two pulse sequences are interlaced in a single acquisition with total of 500 single shots, taking around 5 seconds to execute. 
	\textbf{f}, Change of $f_\text{ESR}$ over the whole randomised benchmarking measurement period. 
	\textbf{g}, Calibration of ESR amplitude, ensuring a fixed $1.75\mu$s~$\pi$-pulse time. Similarly to the process in (e), but now both sequences perform an $X/2$ pulse 32 times followed by an $X/2$ or $-X/2$. 
	\textbf{h}, Change of microwave amplitude, $\Omega_\text{ESR}$ over the whole randomised benchmarking measurement period.
	No strong correlations are observed between all the 4 feedback parameters over 35 hours, suggesting most slow drifts/glitches in the qubit system are independent.
		\hspace*{\fill}
	}
\end{figure*}

We implemented 4 different controllers to ensure the spin qubit environment and parameters do not drift. The first two controllers are responsible for spin-to-charge readout process, and the other two for the Hamiltonian coefficients. \autoref{fig:feedback} shows how all the 4 controllers and their respective parameters change throughout the whole 35 hours of the randomised benchmarking experiment. 
\autoref{fig:feedback}a is the schematic of the circuitry controlling the sensor current $I_\text{sensor}$. The difference between $I_\text{sensor}$ and the desired sensing point $I_\text{ref}$ is passed through a gain of $\beta$, and fed back into $V_\text{TG}$. This controller ensures the sensing signal $I_\text{sensor}$ is always sitting on the most sensitive point for blip detection.
\autoref{fig:feedback}c is the schematic of the circuitry controlling the dark blip count $\text{blip}_\text{dark}$. The dark blip count refers to the excessive blips that occur even when the qubit spin is down, gathered as the blip detection count at the later half of the readout time window. Dark count occurrence is usually caused by not biasing the readout level in the middle of the Zeeman splitting energy. Having the dark blip count being too high or low may cause the readout visibility to become saturated and will have an effect on the analysis of the randomised benchmarking decay rate. Here we set the $\text{blip}_\text{ref}$ to 0.16 for maximum readout visibility for the controller.
The above two controllers are automatically applied by doing extra analysis of $I_\text{sensor}$ traces for each acquisition (a single digitiser data transfer of collective single shot traces of $I_\text{sensor}$), and do not required additional adjustment. The next two controllers require interleaved measurements that are independent of the randomised benchmarking sequence. These are done periodically, after every 16 acquisitions. 
Here, we can modify our Hamiltonian into:

\begin{align}
&H=\Omega_\text{drift}\Omega_\text{ESR}(\Omega_x\sigma_x + \Omega_y\sigma_y) + (f_\text{ESR}+\epsilon_z)\sigma_z,
\end{align}

where $f_\text{ESR}$ is the ESR centre frequency adjustment that can be seen as a multiplier on $\sigma_z$. 
This can cancel the effect of detuning noise offset, $\epsilon_z$.
$\Omega_\text{drift}$ is the effective physical ESR amplitude multiplier that drifts over time and is balanced by $\Omega_\text{ESR}$ through the controller to maintain the relation: of $\Omega_\text{ESR}\Omega_\text{drift}=1$. 
We also have the relation $\Omega_{x,y}' = \Omega_\text{ESR}\Omega_{x,y}$ which is shown in \autoref{fig:device}a
$f_\text{ESR}$ is updated by measuring the difference between the two control sequences, shown in \autoref{fig:feedback}e. We have one sequence of $X/2\Rightarrow{}Y/2$ with 0.2$\mu$s gap, while the other one has the $Y/2$ changed to $-Y/2$. The two calibration sequences would have equal spin up probability that is close to 0.5 if no resonance frequency offset exists and will have a different probability if $f_\text{ESR}+\epsilon_z\ne0$, regardless of other SPAM errors. We then take the spin up probability difference of these two sequences and feed this back into $f_\text{ESR}$ with a certain stable gain, where now the controller will enforce $f_\text{ESR}+\epsilon_z\sim0$, since $\epsilon_z$ has a very slow drift over the calibration period (range of minutes).
Similarly, after calibrating $f_\text{ESR}$, we perform another calibration sequence pair shown in \autoref{fig:feedback}g, where now the first has $X/2$ repeated for 32 times, and followed by another $X/2$ at the end, versus the second having $-X/2$ at the end.
Given that the $f_\text{ESR}+\epsilon_z$ term is negligible at this stage, the spin up probability of these two sequences are also close to 0.5 and only the same when $\Omega_\text{ESR}\Omega_\text{drift}=1$. Any difference of these two probabilities will feedback into $\Omega_\text{ESR}$. The repetition of 32 is chosen for higher accuracy of calibrating $\Omega_\text{ESR}$ whilst still maintaining a stable controller. A repetition number that is higher will give better accuracy but with less tolerance of the drift range. 
This can result same in the same spin up probability where  $\Omega_\text{ESR}\Omega_\text{drift}=A$, $A$ being a number close to 1.
On average 16 acquisitions take around  35 seconds and the two calibrations of $f_\text{ESR}$ and $\Omega_\text{ESR}$ take around 5 seconds each.
Figures~\ref{fig:feedback}~b,d,f,h show a plot of feedback values over the measurement time period for the controllers on their left side. 

During the randomised benchmarking measurement, traces of $V_\text{TG}$,  $V_\text{G1}$, and  $f_\text{ESR}$ appear to be binary/step like, suggesting changes in the qubit environment are more event like rather than drifting. 
The cause of these jump events could include local charge rearrangement, battery switching of gate sources, or local nuclear spin flip. However, $\Omega_\text{ESR}$ appears to be a drift like mechanism, which we believe is due to the high sensitivity of microwave source to temperature and power supply. 
Interestingly, we observe no clear correlation between all the 4 traces; this is a strong indication that $f_\text{ESR}$ jumps of the qubit come from nuclear spin flip, rather than local charge rearrangement that would require a big offset in read-out level $V_\text{G1}$ with given Stark shift level \cite{chan2018}.

\subsection{Randomised benchmarking}

\subsubsection{Theory of randomised benchmarking}

The essence of randomised benchmarking is that it uses long sequences of gates with the aim of amplifying small errors in the implementation of these gates. By choosing the sequences of gates from a unitary 2-design the average noise channel over random sequences of such gates will reduce to a depolarising channel with the same fidelity (to the identity) as the average noise between idealised versions of these gates (the \textit{average fidelity}) \cite{wallman2017}. As here, where the design chosen is the Clifford group, then we call this the \textit{per Clifford fidelity}. The fidelity of a channel to the identity is given by
\begin{equation}
F(\mathcal{E})=\int d\psi \ev{\mathcal{E}\qty(\op{\psi})}{\psi}\,,
\end{equation} 
where the integral is over all pure states $\ket{\psi}$ in accordance with the Haar measure. 

Typically the Clifford group is chosen  because of its importance  as the foundation of many fault-tolerant architectures, meaning that the gates implementing the group are precisely the type of gates likely to appear in such implementations. Whilst randomised benchmarking is a scalable protocol that can be applied to multi-qubit systems, here we are utilizing it to benchmark a single qubit and the discussion below is predicated on the single-qubit system in question.

Typically a randomised benchmarking experiment involves: 
\begin{enumerate}
\item choosing a sequence length $m$, 
\item preparing a state ($\rho$) in the computational basis, 
\item applying the chosen number of random gates, drawn from the unitary 2-design (here the Clifford group),
\item applying a further gate, which if all the previous operations were ideal would return $\rho$ to the computational basis,
\item measuring $\rho$ in the computational basis, to determine if it has been so returned.
\end{enumerate}
The above steps are typically repeated a sufficient number of times to estimate the survival probability for that $m$. A different $m$ is chosen and the whole process repeated to build up an estimate of the survival probablity ($\bar{q}$) over a range of sequence lengths. These estimates are then used to fit to the model: 

\begin{equation}\label{eq:rbformula}
\bar{q}(m)=Ap^m+B\,,
\end{equation}
where $p$ is the depolarising parameter, related to the average fidelity ($\bar{F}$) as: $\bar{F}$ = $(1+p)/2$. Here $A$ and $B$ represent parameters related to the state preparation and measurement (SPAM) errors. 

A variation of the randomised benchmarking protocol allows the fidelity of various Clifford gates to be estimated \cite{Magesan2012}. Here the first protocol is used as a reference run, and the experiment is conducted again, save that a copy of the gate to be estimated is interleaved between the randomly chosen Clifford gates. By making the simplifying assumption that $\bar{F}_\text{combined}=\bar{F}_\text{ref}\bar{F}_\text{C}$ an estimate of the fidelity of that Clifford gate ($\bar{F}_C$) can be obtained. See \cite{Kimmel2014,dugas2016} for a detailed discussion of bounds on the above assumption.

\subsubsection{Randomised benchmarking sequence}
\label{sec:RBM}

\begin{figure*}
	\includegraphics{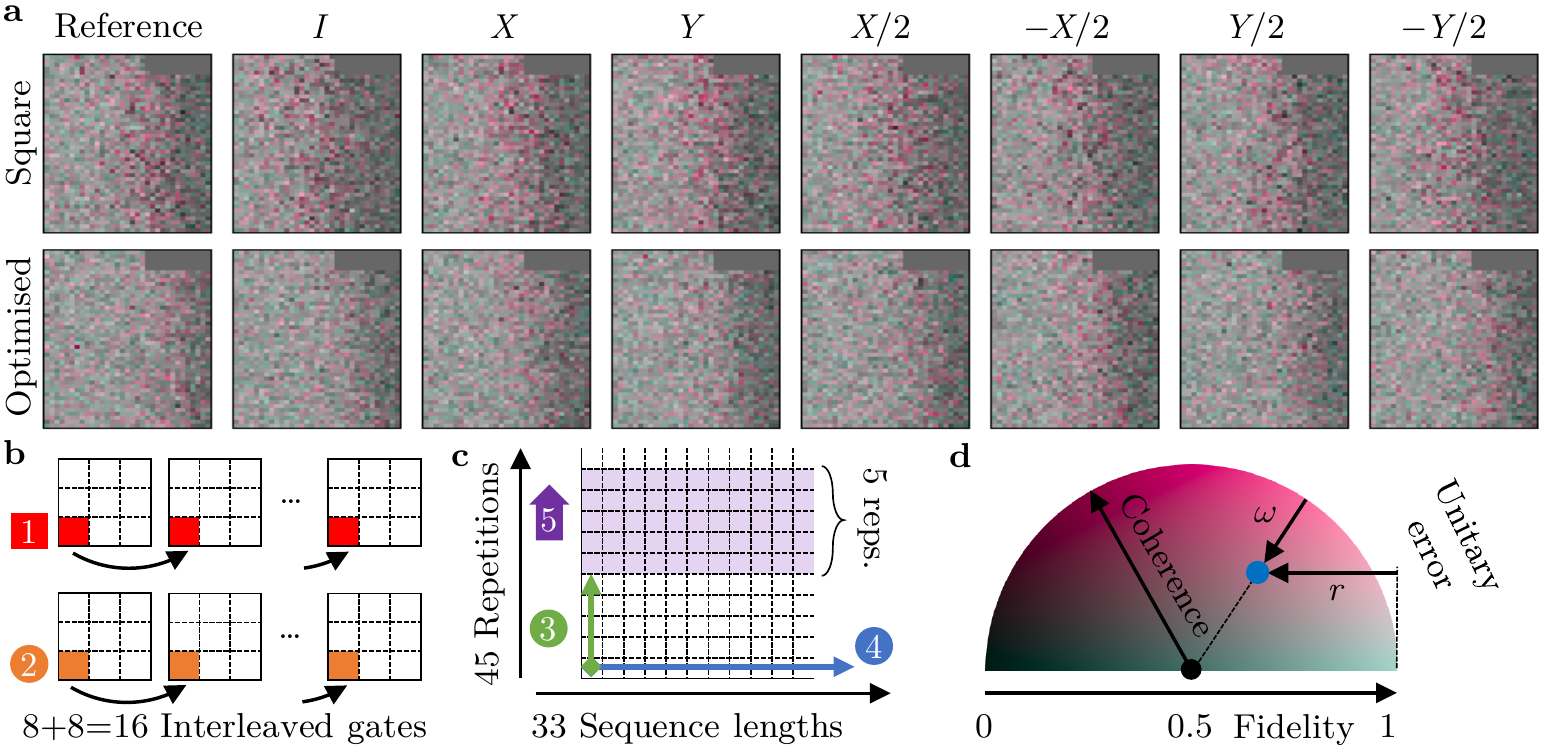}%
	\caption{\label{fig:RBM}
		\textbf{Randomised benchmarking experimental data.}
		\textbf{a}, The realigned partial density matrix data for each acquisition, produced via tomographic readout. The density matrix is realigned to spin up with respect to its expected final state ($\pm{}XY\mkern-2mu Z$), and the phase information is removed. The partial density matrix is colour encoded via a semi-circle (d).
		\textbf{b}, The sequence of randomised benchmarking experiments. Each grid box represents the collection of density matrices for a single interleaved gate, formatted as in (a). 
		\textcircled{1}: Step through each interleaved gate sequence using standard square pulses, Reference, $I$, $X$, $Y$,$X/2$,$-X/2$,$Y/2$,$-Y/2$, for a total of 8 acquisitions. 
		\textcircled{2}: Similar to \textcircled{1} but with GRAPE optimised pulses, for a total of 8 acquisitions.
		\textbf{c}, Sequence of randomised benchmarking experiments after stepping through each interleaved gate sequence.
		\textcircled{3}: Repeat the $8+8=16$ acquisitions 5 times, stacked up on the $y$-axis. 
		\textcircled{4}: Step through sequence lengths, $m$ of [1, 2, 3, 4, 5, 6, 8, 10, 13, 16, 20, 25, 32, 40, 50, 63, 79, 100, 126, 158, 200, 251, 316, 398, 501, 631, 794, 1000, 1259, 1585, 1995, 2512, 3162], stacked up on the $x$-axis, total of 33 steps. 
		\textcircled{5}: repeat everything again, stacked up on the $y$-axis, total of 9 repetitions. 
		\textbf{d}, Colour semi-circle representation of a partial (phase-less) density matrix. Higher lightness means higher fidelity, and higher red component means higher $XY$ spin component mixed in. 
		\hspace*{\fill}
	}
\end{figure*}

We can now perform the randomised benchmarking experiment using the methods described earlier. The results are shown in the main text (Figure.~\ref{fig:sequences}) and in Figure.~\ref{fig:RBM}.
We also present the data as follows: for every measurement acquisition of a randomised benchmarking sequence, we obtain a density matrix which is reconstructed via 120 single shot spin readouts with tomographic measurement (see main text). 
The density matrix can be rotated in a way that its expected final state would have aligned to spin up ($+Z$), followed by removing the $XY$ phase angle while maintaining its magnitude.
This produces a realigned partial density matrix map which is colour encoded in \autoref{fig:RBM}(a) as per the  colour-semi-circle in \autoref{fig:RBM}(d). The maps are grouped in different interleaved gates and contain the complete measurement data set for every single acquisition that is studied in this paper, before any averaging and analysis take place.
In order to present the measurement data in as raw a form as possible, other than the realigned phase information being taken away (since it only has trivial physical meaning in a randomised benchmarking experiment), no other corrections including SPAM error re-normalisation are performed.
The colour point that has higher brightness means the measured final state from a randomised benchmarking sequence has higher fidelity.
If the colour red is mixed in the data point, this suggests unitary errors occurred which may result in a measurement having low fidelity but a high coherence/unitarity (analysis in \autoref{fig:sequences}).
Note that there is no colour saturation in \autoref{fig:RBM}(a), meaning there are no data compression losses unless through limitation of viewing/printing device for this paper. However, the colour-semi-circle at high coherence/visibility is saturated, but no experimental data points lie within those regions. The grey boxes at the top right corner for each map are unperformed data points due to early termination of the measurement.

\autoref{fig:RBM}(b,c) describes how the whole randomised benchmarking experiment is stepped through in time sequence, and
the numbers in circle are the order of stepping in time.
To begin with, note that the Clifford gate sequences in every single data point shown in \autoref{fig:RBM}(a) are re-randomised and different.
Now, we have \textcircled{1} (square) and \textcircled{2}(optimised) that steps through the different interleaved gates in \autoref{fig:RBM}(b).
It starts from the standard square pulse reference (no interleaved gate) with a randomised Clifford gate sequence. Once tomographic readout acquisition is done, it moves on to the next interleaved gate, $I$, and regenerates a new randomised Clifford gate sequence with same sequence length, $m$, which takes about 1.7 seconds (at short $m$). After the last interleaved gate,$-Y/2$ acquisition is completed, which concludes process \textcircled{1}, the same measurement is repeated again but with the GRAPE optimised pulses, referred to as process \textcircled{2}. At the end of $\textcircled{1}+\textcircled{2}$, the frequency and power calibration then kicks in to adjust the qubit environment (see above). A total of 16 acquisitions are cycled through (8 interleaved gates and two types of Clifford gate pulses) and this takes about 40 seconds (at short $m$) including the calibration.
When the interleaved gate cycle is done, we now move to \autoref{fig:RBM}(c) where process \textcircled{3} starts.
Process \textcircled{3} is a simple 5 repetition sequence of $\textcircled{1}+\textcircled{2}+\text{calibration}$, it is repeated on the $y$-axis in \autoref{fig:RBM}(a), and takes about 3 minutes (at short $m$) to complete. Process \textcircled{4} changes $m$ after completion of process \textcircled{3}, stepping through [1, 2, 3, 4, 5, 6, 8, 10, 13, 16, 20, 25, 32, 40, 50, 63, 79, 100, 126, 158, 200, 251, 316, 398, 501, 631, 794, 1000, 1259, 1585, 1995, 2512, 3162] sequentially, a total of 33 steps shown on the $x$-axis of \autoref{fig:RBM}(a), and takes about 250 minutes to complete. Finally, \textcircled{5} repeats everything above a total of 9 times and stacks up on the $y$-axis of \autoref{fig:RBM}(a), with a final product of 45 rows. The complete measurement can be expressed as process stack $(\textcircled{1}+\textcircled{2}+\text{calibration})\times\textcircled{3}\times\textcircled{4}\times\textcircled{5}$, and lasts for 35 hours.

\subsubsection{Eliminating the nuisance parameter \texorpdfstring{$B$}{B}.}

We note that using tomographic measurements also allows a variation of the RB protocol similar to variations previously discussed in the literature \cite{Knill2008,fogarty2015,Muhonen2015}. For any particular sequence the tomographic measurements at the end of the sequence include not only a measurement that corresponds to the expected `maximal-overlap' measurement of the state, but also one that corresponds to a `minimal-overlap' measurement. This `minimal-overlap' measurement can be included by setting $\bar{q}(m,s)=1-\bar{q}(m,s)$ for each such measurement and combining this into in the average estimate of the survival probability for each sequence length, $m$. If this is done the constant $B$ gets mapped to $[B+(1-B)]/2=1/2$. This removal of the SPAM parameter $B$ leaves only two free parameters with which to fit the data, leading to tighter credible regions for the parameter of interest ($p$).

The randomised benchmarking procedure described above was carried out for 33 different sequence lengths of $m$ (see \autoref{fig:sequences}). The survival percentage for each sequence of a particular length was averaged (as discussed above) and a weighted least squares non-linear fit was performed to the data, using \autoref{eq:rbformula}, with $B$ set to 0.5. The data points were weighted by the inverse variance of the observed data at a particular $m$.

To take into account possible gate dependent noise, the non-linear fit to the data was re-analysed, this time ignoring the $m$ of less than four (which are the only $m$ that are likely to be noticeably affected by gate-dependent noise)~\cite{wallman2017}, with no significant impact on the results.  To finalise the analysis, Qinfer \cite{Granade2014,qinfer-1_0} was used to analyse the data using Bayesian techniques (a sequential Monte-Carlo estimation) of the parameter~$p$. As can be seen in \autoref{fig:sequences}b the credible region found accords with the least square fit methods. This provides an indication as to correctness of the model, which might not be the case if the system were still impacted by low frequence noise~\cite{Ball2016}. Finally we note that the use of repeat sequences complicates the analysis surrounding the use of least squares estimates and the Bayesian techniques used by Qinfer. However, using bootstrapping methods on the data confirms the robustness of the estimates. 

\subsubsection{Determining the unitarity from tomographic measurements}\label{sec:tm_unitarity}

The tomographic measurements allow the unitarity of the average noise channel to be measured \cite{Wallman2015a}. The protocol is similar to a randomised benchmarking experiment, save that no inverting gate is applied and the resulting state is best measured as an average over the non-identity Pauli operators, known as the \textit{purity measurement}. For a single qubit this can be accomplished by measuring $\mathcal{Q}= \langle S_x\rangle^2+\langle S_y\rangle^2 + \langle S_z \rangle^2$, where each expectation value is taken with respect to the state in question. The projective measurements carried out by the tomography allow us to make numerical estimates for each of the components of the purity measurement and thus for $\mathcal{Q}$.  Then using the techniques discussed above, this is fit to a curve of the form 
$\mathcal{Q}(m)=A+Bu(\mathcal{E})^{(m - 1)}
$, where $u(\mathcal{E})$ is the unitarity and $A,B$ are parameters that absorb SPAM noise.

\bibliography{library}
%

\end{document}